\documentclass[showpacs,twocolumn,prl,aps,english,superscriptaddress]{revtex4}
\usepackage{amsmath}
\usepackage{graphicx}
\usepackage{amssymb}
\usepackage{babel}
\usepackage{color}

\newcommand{\rr}{{\bf r}}

\newcommand{\R}{{\bf R}}
\newcommand{\K}{{\bf K}}

\newcommand{\av}[1]{\langle{#1}\rangle}

\newcommand{\N}{{\frak N}}
\newcommand{\F}{{\frak F}}

\begin{document}

\author{Dmitry Solenov}
\email{d.solenov@gmail.com}
\altaffiliation{Present address: Naval Research Laboratory, 4555 Overlook ave., SW Washington, District of Columbia 20375, USA}
\affiliation{National Research Council, National Academies, Washington, District of Columbia 20001, USA}

\author{Chad Junkermeier}
\altaffiliation{Present address: Naval Research Laboratory, 4555 Overlook ave., SW Washington, District of Columbia 20375, USA}
\affiliation{National Research Council, National Academies, Washington, District of Columbia 20001, USA}

\author{Thomas L. Reinecke}
\affiliation{Naval Research Laboratory, Washington, District of Columbia 20375, USA}

\author{Kirill A. Velizhanin}
\email{kirill@lanl.gov}
\affiliation{Theoretical Division, Los Alamos National Laboratory, Los Alamos, NM 87545, USA}

\title{Tunable Adsorbate-Adsorbate Interactions on Graphene}

\begin{abstract}
We propose a mechanism to control the interaction between adsorbates on graphene. The interaction between a pair of adsorbates---the change in adsorption energy of one adsorbate in the presence of another---is dominated by the interaction mediated by graphene's $\pi$-electrons and has two distinct regimes. {\it Ab initio} density functional, numerical tight-binding, and analytical calculations are used to develop the theory. We demonstrate that the interaction can be tuned in a wide range by adjusting the adsorbate-graphene bonding or the chemical potential.
\end{abstract}


\maketitle

Graphene is a one-atom-thick sheet of carbon. It is of interest for a variety of applications including transport, photovoltaics, and DNA manipulations \cite{Geim2007-183,DasSarma2011-407,Wehling2008-173}. The unique electronic transport properties of graphene stem from its two-dimensional honeycomb carbon network structure. However, wider applications are limited by difficulties in opening a bandgap and its lack of processability. 
Chemical modifications can address both limitations via changing the electronic, chemical, and mechanical properties of graphene \cite{Englert2011-279,Boukhvalov2009-4373,Kozlov2011-2638,Rotenberg2012}. As a result, design and control of functionalization has become an important challenge.

Atoms or molecules can adsorb on graphene via van der Waals interactions (physisorption) or by forming chemical bonds with one or several carbon atoms on the surface (chemisorption). Functionalization of graphene is typically based on chemisorbed atoms or molecules that remain chemically active or provide other functions, e.g., influence conductance. Adsorbate-graphene interactions for various types of adsorbates have been discussed  \cite{Robinson,Wehling2009-085428,Wehling2010-056802,Liu2011-235446,Krasheninnikov2011-625,Lambin2012-045448,Wehling2009-125,Suarez2011-146802,Solenov2012-095504,Shih2012-8579}. Recently, it has been recognized that the adsorption energy of an atom or molecule can depend on other adsorbates. This
can become especially important at large adsorbate concentrations \cite{Shytov2009-016806,Cheianov2009-233409,Cheianov2009-1499,Kopylov2011-201401,Huang2012-125433}. Therefore, understanding the nature of the adsorbate-adsorbate (A-A) interactions is a key to the design and control of functionalization of graphene.

In this Letter we investigate the microscopic mechanisms underlying the interaction between two atoms or molecules chemisorbed on graphene at a distance from each other. We demonstrate that A-A interaction is dominated by an exchange interaction mediated by graphene's $\pi$-electrons and has two distinct regimes. The change of the interaction energy as a function of distance between adsorbates involves three phases: (1) change of sign associated with graphene sublattice, (2) variations due to the momentum difference between the two Dirac cones of graphene, (3) change of sign due to change in the scattering mechanism. The first two phases are enforced by graphene's lattice \cite{Shytov2009-016806} and are also present in magnetic RKKY interaction in graphene \cite{Sherafati,Saremi}. The third phase appears due to significant restructuring of the electron-mediated interaction and has not been noticed to date. This phase, $\Theta(\R)$, changes dramatically only once at some critical distance $R_C$ between the two adsorbates. The change takes place on the scale of a single carbon-carbon bond resulting in an abrupt sign reversal. The critical distance $R_C$ depends on the bonding mechanism between each adsorbate and graphene as well as the chemical potential. As a result, A-A interaction can be controlled in a wide range by adjusting $R_C$. We use a combination of first-principles density functional theory (DFT) calculations \cite{Giannozzi09395502}, numerical tight-binding (TB), and analytical functional integral approaches to identify the role of different microscopic mechanisms responsible for the interaction. We illustrate the results with fluorine (F) and amine (NH$_2$) adsorbates. 

Chemisorbed atoms or molecules can form one (monovalent) or several (multivalent) bonds with the graphene carbon (C) atoms.
Monovalent adsorbates, including F and NH$_2$, adsorb on top of one of the C atoms forming (typically) a mixed covalent-ionic bond. For example, in the case of F the covalent component of the bond is formed by the overlap of F $pz$ orbital and $s$-$pz$ hybridized orbital of C. The C atom participating in this bond changes its hybridization from $sp^2$ to $sp^3$ taking one $pz$ electron out of the $\pi$-system. This leads to creation of more complex states near each adsorbate site as well as local deformation of graphene. These modifications can lead to four possible A-A interaction mechanisms: (i) direct overlap of adsorbate's electron orbitals; (ii) gain or loss of energy due to change of elastic deformation energy of the graphene; (iii) Coulomb (or electrostatic) interaction between charges on each adsorbate; and (iv) interaction induced by multiple scattering of graphene's $\pi$-electrons off the adsorption sites. The first contribution is effective only for atoms adsorbed in direct proximity of each other \cite{Nguyen}. Lattice deformation produces $1/R^3$ contribution \cite{Lau,Peyla} that is weak $\sim 1-10$ meV, except at the shortest distances. In what follows we demonstrate that the primary contribution to A-A interaction at larger distances comes from the interplay between Coulomb and $\pi$-electron scattering-induced interactions.

The Coulomb and electron exchange interactions in the system of two adsorbates on graphene can be described by a TB Hamiltonian \cite{comment-TB}
\begin{eqnarray}\nonumber
&H&\!\!\! =\!-\!\!\!
\sum_{\av{ij}'}\!
tc^\dag_ic_j
\!+\!\!\!\!\!\!
\sum_{n=1,2;s}\!\!\!\!
u_{n,s} a^\dag_{n,s} a_{n,s}
\!-\!\!\!\!\sum_{\av{in},s}\!\!t'_{n,s}(c^\dag_ia_{n,s}\!\!+\!a_{n,s}^\dag c_i) 
\\\label{eq:H}
&+&
\!\!\!\!\!\!
\sum_{i\neq j}\!
U_{i,j}\hat n_{i}\hat n_{j}
\!\!
+
\!\!\!\!\!\!\!\!\!
\sum_{i,n=1,2;s}
\!\!\!\!\!\!
U^{(s)}_{i,n}\hat n_{i}\hat {\frak n}_{n,s}
\!\!
+
\!\!\!
\sum_{s,s'}
U^{(s,s')}_{1,2} \hat {\frak n}_{1,s}\hat {\frak n}_{2,s'} 
\!
+
\!\!
H_{I}
\!.
\end{eqnarray}
Here the first term describes graphene's $\pi$-electrons, the $c^\dag_i$ ($c_j$) are creation (annihilation) operators of $pz$ states on C atoms, and $t$ is the hopping integral between the nearest $pz$ states. The two adsorbates ($n=1,2$) can form chemical bonds with several nearby C atoms removing their $pz$ states from the graphene's $\pi$-system, as indicated by the prime in $\av{ij}'$. The TB model is formulated by first considering each adsorbate together with C's to which it is bonded as an isolated adsorbate-carbon complex (ACC).
The states on ACC that are accessible to itinerant $\pi$-electrons, $a_{n,s}^\dag$ ($a_{n,s}$), are not single electron states. They are formed as a result of overlap of atomic orbitals and strong Coulomb interaction between multiple electrons occupying these orbitals. The ACC is then connected (by tunneling and Coulomb interaction) to the nearby graphene's $pz$ orbitals. The hopping integrals between ACC states and $pz$ states on other C atoms are denoted by $t'_{n,s}$ in the third term. This approximation is justified by the fact that the Coulomb interaction and tunneling within ACC is stronger.
Formation of these states is, in general, a complex problem that goes beyond the scope of the present Letter, and we extract $u_{n,s}$ and $t'_{n,s}$ from DFT calculations.
The fourth term in (\ref{eq:H}) is the Coulomb interaction between the $\pi$-electrons ($\hat n=c^\dag c$). The fifth term represents the Coulomb interaction between the $\pi$-electrons and electrons that occupy ACC states ($\hat {\frak n}=a^\dag a$). The last two terms encode the Coulomb interaction between electronic densities and ionic cores of {\it different} ACCs. 

\begin{figure}
\centering{}
\includegraphics[width=0.99\columnwidth]{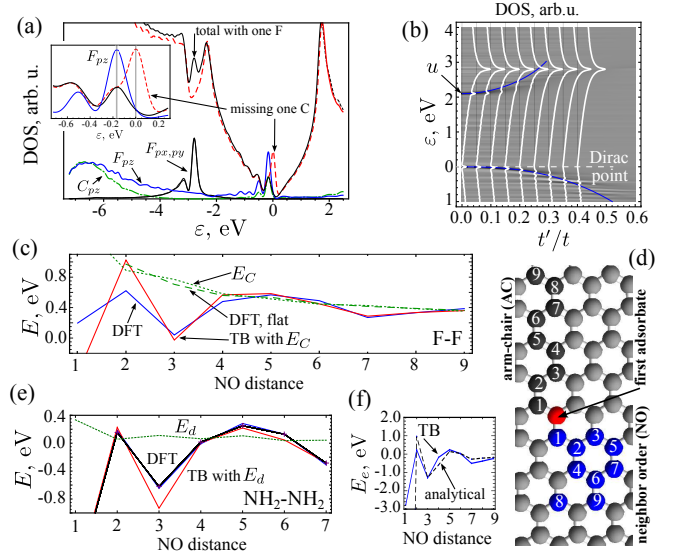}
\caption{\label{fig1}
The results of DFT, TB, and analytical calculations. 
(a) DFT DOS (and PDOS) for the system of graphene with a {\it single} adsorbed F. The inset shows the shift of the DOS peak due to nonzero $t'$. The (spin-restricted) DOS of graphene with one C removed is shown for reference. 
(b) TB DOS as a function of $t'$. The two solid (blue) curves are the analytical solutions, as discussed in the text. The white curves are total DOS. The background density plot is the difference between TB DOS with and without F atom: the two main peaks (darker shade) follow the analytical solution for small-to-intermediate values of $t'/t$.
(c) Interaction energy between two F's as a function of neighbor order (NO) distance as shown in panel (d). (d) Numbering convention for the distance between the two adsorbates: one adsorbate is at position ``0" (red) and the other at any of the other marked sites. (e) Interaction energy between two NH$_2$'s (crosses show spin-unrestricted result). (f) Comparison between numerical TB and analytical estimates to the interaction energy. 
}
\end{figure}

The parameters $u$ and $t'$ of the Hamiltonian (\ref{eq:H}) can be obtained from  DFT calculations \cite{DFT-details,Giannozzi09395502,Grimme061787,Barone09934,Grimme11211} by analyzing either the electronic spectrum or the density of states (DOS) of graphene with a {\it single} adsorbate. We consider F as an example to illustrate this parametrization.
As we see in Fig.~\ref{fig1}(a), the attachment of F creates a peak in the DOS near the Dirac point. This peak, however, does not come from ACC. It is a boundary defect (or Tamm) state (TS) formed as the result of excluding one carbon's $pz$ orbital from the $\pi$-system \cite{com-tamm}. 
This peak is shifted away from the Dirac point due to interaction (tunneling) with the nearest ACC states. In the case of F it is sufficient to consider only one ACC state that is nearest in energy (NS) and that can be identified approximately as a C-F anti-bonding [see projected DOS (PDOS) in Fig.~\ref{fig1}(a)]. To extract the values of $u$ and $t'$ we compute the TB DOS \cite{TB-details} for one adsorbate using the first three terms of (\ref{eq:H}). The resulting DOS [see Fig.~\ref{fig1}(b)] has two peaks associated with the adsorbate that are used to fit the DFT data. We obtain $t'\approx 0.27t$ and $u\approx 0.75t$. The shift of these peaks due to tunneling between TS and NS can also be obtained analytically in the limit of linear dispersion relation as $\varepsilon(\varepsilon-u) = \pi\sqrt{3}t'^2/2\log|3^{1/4}\sqrt{\pi}t/\varepsilon|$. For small $t'$, $\varepsilon \approx u\pm\sqrt{u^2+(3t'/2)^2}/2$. The parametrization for other types of adsorbates can be done similarly, e.g., for NH$_2$ we obtain $t'\approx 0.25t$ and $u\approx 1.11t$ \cite{com-mag-shifts,Santos}.

\begin{figure}
\centering{}
\includegraphics[width=0.99\columnwidth]{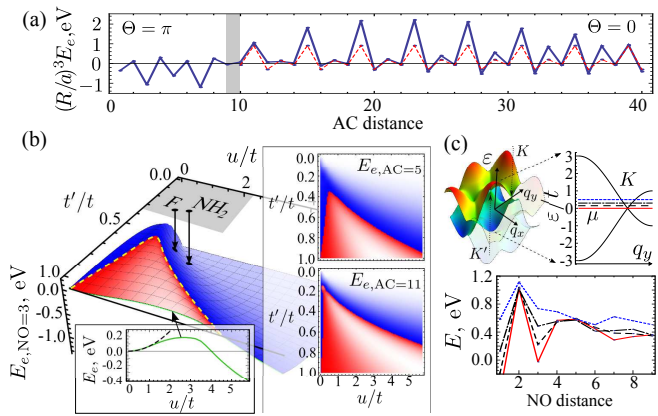}
\caption{\label{fig2}
Adsorbate-adsorbate interaction energy for different scattering strengths and distances. (a) Transition from strong to weak scattering regimes with distance: at short distances the minima occur when adsorbates are on different sublattices [odd sites along AC in Fig.~\ref{fig1}(d)]; at $R_C$ of about AC=10 and above the phase changes, and the minima occur when adsorbates are at the same sub-lattice sites (even sites along AC). The dashed line is the weak coupling limit. (b) The A-A interaction as a function of $t'$ and $u$ at NO=3 (and AC=3) distance between the adsorbates. The dashed curve represent the transition boundary ($R_C$). The values corresponding to F and NH$_2$ are marked (approximately) on the shaded plane. The bottom inset shows $E_e(u)$ at $t'=t$ (solid curve). It starts quadratically in $u$ (dashed curve). The right inset shows interaction energy for AC=5 and AC=11. In all surface plots red corresponds to positive values and blue to negative (the lighter shade reflects larger numbers). (c) The A-A interaction energy between two F's as a function of distance for different chemical potentials $\mu/t = 0,0.15,0.3,0.5$: solid (red), dashed, dash-dotted, and dotted (blue) lines, respectively. The top subpanel shows the values of $\mu$ relative to graphene's spectrum.
}
\end{figure}

The A-A interaction energy, $E(\R)$, between {\it two} adsorbates a distance $\R=\R_2-\R_1$ apart involves two components: electrostatic interaction energy $E_C$ and $\pi$-electron scattering-induced interaction energy $E_e(\R)$. In the case of a monoatomic adsorbate, such as F, $E_C$ can be approximated well by $e^2Z^2/4\pi\varepsilon_0 r$, where $Z$ is the effective charge drawn to the adsorbate complex obtained from DFT ($Z\approx -0.39$ for F). Multipole components can become significant for more complex adsorbates. For example, in the case of NH$_2$, the net charge on the adsorbate molecule is small and the electrostatic interaction is dominated by the dipole-dipole interaction [see Fig.~\ref{fig1}(e)]. 

The screening [due to the fourth and the fifth terms in (\ref{eq:H})] does not alter the electrostatic interaction significantly at distances of interest. This can be seen either from the interaction energy, as we will see shortly, or from the DFT calculations, e.g., by restricting C atoms to their original (clean, flat graphene) positions [see ``DFT, flat" curve in Fig.~\ref{fig1}(c)]. This restriction enforces $sp^2$ hybridization on the C atoms under adsorbates and eliminates elastic deformation of the graphene sheet. In the case of F, it makes the C-F bond essentially ionic. As a result, the electron scattering due to the second and the third terms in (\ref{eq:H}) is attenuated. By comparing $E_C(\R)$ and ``DFT, flat" results in Fig.~\ref{fig1}(c) we conclude that the fourth and fifth terms in (\ref{eq:H}) do not contribute significantly to the A-A interaction. To further demonstrate the dominant role of the first three terms in (\ref{eq:H}) we compare $E(\R)$ computed with fully relaxed DFT to that from numerical TB \cite{TB-details} [together with $E_C(\R)$]. The energies for two F and two NH$_2$ are shown in Fig.~\ref{fig1}(c) and (e) respectively. The DFT and TB curves show remarkable agreement at distances starting from about NO=3 [Neighbor Order distance, see Fig.~\ref{fig1}(d)]. Larger deviations at short distances in both cases can be attributed to a combination of local deformation, direct overlap of adsorbate orbitals, and the next-nearest neighbor hopping, which are not included in (\ref{eq:H}). Note that the modification of the interaction energy due to possible formation of magnetic moments \cite{Santos} is several orders of magnitude smaller compared to the magnitude of the A-A interaction energy, cf. the NH$_2$ DFT results for spin-restricted and spin-unrestricted DFT in Fig.~\ref{fig1}(e).

The A-A interaction energy, $E_e(\R)$, oscillates depending on whether adsorbates are on the same or different sublattices \cite{Geim2007-183}. In addition, the interaction varies on the scale of $1/|{\bf K}-{\bf K}'|$ in a fashion similar to the RKKY interaction between magnetic impurities on graphene \cite{Sherafati}. There is an additional phase factor, $\Theta$, however, that depends on the values of $u$ and $t'$ and causes a dramatic change in the interaction at some distance $R_C$. This can be most easily seen in the numerical TB results. Figure~\ref{fig2}(a) shows the interaction energy vs. distance between adsorbates placed at AC sites [Arm-Chair sites, see Fig.~\ref{fig1}(d)]. At shorter distance adsorbates attract when on different sublattices. At large distance attraction takes place when adsorabates are on the same sublattice. The transition between these two regimes happens abruptly at $R_C$ on the scale of a single C-C bond length [see shaded area in Fig.~\ref{fig2}(a)]. The location of $R_C$ varies as a function of both $u$ and $t'$. The overall amplitude of the interaction crosses over slowly, approaching $1/R^3$ as $R\to\infty$.
In order to track these changes we analyze the interaction energy at specific sites [NO=3, AC=3, AC=5, and AC=11 in Fig.~\ref{fig2}(b)] as a function of $u$ and $t'$. At large $u$ and/or small $t'$ the interaction approaches a constant value and is attractive when the adsorbates are on different sublattices. We will denote this as a {\it strong scattering} regime (the limits $u=0$, $t'\to 0$ and $t'=t$, $u\to\infty$ are equivalent). For smaller values of $u$ and larger $t'$ the sign of the interaction changes indicating a different, {\it weak scattering},  regime. This change occurs at different values of $u$ and $t'$ for different distances between the adsorbates with $R_C$ going to infinity when $u\to\infty$ or $t'\to 0$ [see, Fig.~\ref{fig2}(b)].

The values of $u$ and $t'$ represent the bond between each adsorbate and graphene.
As a result, changes in the bonding characteristics, e.g., due to changes in internal structure of an adsorbed molecule, can shift $R_C$ leading to dramatic modification of the A-A interaction at different distances. This is an interesting fundamental effect rooted in the physics of adsorbate-graphene interaction, and it also provides the control necessary for a variety of applications involving, e.g., surface molecular transport and assembly \cite{Solenov2012-095504}.

To understand $E_e(\R)$ and $\Theta$ we examine an analytical solution for the two limiting cases: (i) the {\it weak scattering} case, small $u$, and {\it strong scattering} case, $u\to\infty$. In both cases we set $t'=t$ for definiteness and work at zero chemical potential. The value of $E_e(\R)$ can be obtained from the electron free energy using the linked cluster expansion \cite{Mahan}. The expression for $E_e(\R)$ can be rearranged to $E_e(\R)\!=\!2\!\int \!\!d\rr d\rr'V_1(\rr)V_2(\rr')J(\R\!+\!\rr\!-\!\rr')$,
\begin{eqnarray}\label{eq:Jij} 
J(\rr_{ij})\equiv J_{ij}(\rr_{ij})=\int \frac{d\omega}{2\pi}G^{(0)}_{ij}(i\omega,\rr_{ij})G_{ji}(i\omega,\rr_j\rr_i).
\end{eqnarray}
Here $V_{1(2)}$ is the scattering potential due to the second and third terms of Hamiltonian (\ref{eq:H}), $\rr_{ij}=\rr_i-\rr_j$ is the distance between two lattice sites, one at sublattice $i$ and the other at $j$. The function $G_{ij}^{(0)}(i\omega,\rr)$ is the bare equilibrium Green's function \cite{Sherafati} and $G_{ij}(i\omega,\rr_i\rr_j)$ is the dressed Green's function, which contains multiple scattering events \cite{Mahan}. In the case $t'=t$, $V_n(\rr) = u V_C\delta(\rr-\R_n)$ ($V_C$ is the volume of the primitive cell). The function $G(i\omega,\R_1\R_2)$ can be found exactly:
\begin{equation}\label{eq:fullG} 
G(i\omega,\R_1\R_2) = \N(i\omega,\F(i\omega,\R))G^0(i\omega,\R)\N(i\omega,0),
\end{equation}
where $\N(i\omega,X) = [1- u V_C G^0(i\omega,0)/\sqrt{2}-X]^{-1}$ and $\F(i\omega,\R) = (uV_C)^2G^0(i\omega,\R)\N(i\omega,0)G^0(i\omega,-\R)/2$. Matrix products are implied where $ij$ indexes are dropped.

When the scattering is weak, $u\to 0$, $G \to G^{(0)}$ and we obtain 
$E_e(\R)= 2 u^2 V_C^2 J^{\rm RKKY}_{ij}(\R)$, recovering the standard RKKY range function for graphene \cite{Sherafati}
\begin{equation}\label{eq:JRKKY} 
J_{ij}^{\rm RKKY}\!(\R)\!=\!\!\frac{(\!-\!1)^{i-j+\!1}\!\sqrt{3}}{3^{i-j}2^6\pi t a^4}
\frac{1\!\!+\!\cos[\R(\K\!\!-\!\K'\!)\!+\!(2\theta\!+\!\pi)\delta_{i\neq j}]}{(R/a)^3}\!,
\end{equation}
where $a$ is the lattice parameter. The A-A interaction energy obtained numerically approaches this result at $R>R_C$ [see Fig.~\ref{fig2}(a), dashed curve].

In the limit $u\to\infty$, the frequency integral in Eq.~(\ref{eq:Jij}) is given predominantly by $\omega\sim at/R$. As a result, at large (but finite) distances we can use $\N(i\omega,\F(i\omega,\R))\N(i\omega,0) \sim -1/\F(i\omega,\R)$. Note that the limits $R\to\infty$ and $u\to\infty$ (or $t'\to 0$) do not commute. We obtain \cite{com-alt-JAA}
\begin{equation}\label{eq:StrongJ1} 
J_{ij}(\R)\!\approx\!\frac{(\!-3)^{i-j}\pi\sqrt{3}t}{2^6 (R/a)}
\frac{1+\cos\R(\K-\K')}{1\!+\!\cos[\R(\K\!\!-\!\!\K')\!+\!(2\theta\!+\!\pi)\delta_{ij}]}.
\end{equation}
The $E_e(\R)$ given by (\ref{eq:StrongJ1}) is shown in Fig.~\ref{fig1}(f) with the numerical TB results. Deviations of the analytical result from numerical at short distances are due to approximate integration and the use of linear dispersion. The renormalization factor $\N(i\omega,\F(i\omega,\R))$ can also provide an analytical estimate for $R_C$, and we find $R_C/a\sim u/t$ (when $t'=t$).

From the above analytical results it is also evident that $R_C$ will depend on the chemical potential, $\mu$. Changes due nonzero $\mu$ in $G^{(0)}G^{(0)}$ are not significant up to the point when the scattering processes are no longer confined to well-defined Dirac cones [see the top subpanel in Fig.~\ref{fig2}(c)]. The factors $\N$, however, are more sensitive and can change substantially as a function of $\mu$. As an example, we consider the system of two F adsorbates. In Fig.~\ref{fig2}(c) we plot $E(\R)$ for several values of $\mu$ that can be achieved by backgating or chemical doping. Significant changes in the A-A interaction can be seen even for relatively small changes in $\mu$.

To summarize, we have demonstrated that the interaction between adsorbates on graphene is a function of adsorbate-graphene bonding. The interaction has two distinct regimes at longer ($\sim 1/R^3$) and shorter ($\sim 1/R$) distances. The transition between these two regimes results in change of the phase in the oscillatory behavior of the interaction. It takes place at the length scale of a single C-C bond and can shift as a function of adsorbate-graphene bonding parameters. This provides a novel mechanism to control and engineer adsorbate-adsorbate interaction on graphene for a variety of applications. The transition between two distinct interaction regimes can also be helpful in understanding the distribution of adsorbates in graphene. It can be particularly important in systems where direct Coulomb $1/R$ repulsion or attraction is not present (e.g., when an adsorbate induces a higher-order multipole field, as in the case of NH$_2$) or is effectively screened (e.g., due to high carrier density).

This work was supported in part by the ONR, NRC/NRL, and by DOE at LANL under Contract No. DE-AC52-06NA25396. Computer resources were provided by the DOD HPCMP.

\end{document}